\documentstyle[12pt]{article} 

\font\tenrm=cmr10 

\newcommand{\bref}[1]{(\ref{#1})} 
\newcommand{\ct}[1]{\cite{#1}}

\newcommand{\be}{\begin{equation}} 
\newcommand{\ee}{\end{equation}} 

\def\theequation{\thesection.\arabic{equation}}
\def\@eqnnum{{\rm (\theequation)}}

\def\lsim{\mathrel{\rlap{\lower4pt\hbox{\hskip1pt$\sim$}}
    \raise1pt\hbox{$<$}}}
\def\gsim{\mathrel{\rlap{\lower4pt\hbox{\hskip1pt$\sim$}}
    \raise1pt\hbox{$>$}}}
\def\frac#1#2{{{#1} \over{#2}}} 
\def\ul{\underline}

\begin{document}  
\begin{titlepage} 
  
\begin{flushright}  
{CU-TP-948 \\ }
{hep-ph/9910559 \\ }  
{\hfill October 2000 \\ }  
\end{flushright}  
\vglue 0.2cm  
	   
\begin{center}   
{ 
{The Standard Model in Its Other Phase \\ }  
\vglue 1.0cm  
{Stuart Samuel $^{1}$ \\ }   
\vglue 0.5cm  

{\it Department of Physics\\}
{\it Columbia University\\}
{\it New York, NY 10027, USA\\} 
\vglue 0.4cm 
and
\vglue 0.4cm  
{\it Department of Physics\\}
{\it City College of New York\\}
{\it New York, NY 10031, USA\\} 

\vglue 0.8cm

%\vglue 0.3cm  
  
{\bf Abstract} 
} 
\end{center}  
{\rightskip=3pc\leftskip=3pc 
%\noindent  
\quad The standard model of particle physics 
is analyzed for the case of a Higgs potential 
not favoring spontaneous electroweak symmetry breaking 
to gain  
insight into the physics of the standard model. 
Electroweak breaking still takes place, 
and quarks and leptons still acquire masses but 
through bosonic technicolor. 
This ``other'' phase of the standard model 
exhibits interesting phenomena.  
}

\vfill

\textwidth 6.5truein
\hrule width 5.cm
\vskip 0.3truecm 
{\tenrm{
\noindent 
%$^*$Permanent address: Department of Physics, 
%City College of New York,\\
%\hspace*{0.2cm}New York, NY 10031, USA.\\
%$^*$Permanent address.\\   
$^{1)}$\hspace*{0.15cm}E-mail address: samuel@cuphyc.phys.columbia.edu \\ }}
 
\eject 
\end{titlepage}

\newpage  

\baselineskip=20pt  

{\bf\large\noindent I.\ Introduction}\vglue 0.2cm
\setcounter{section}{1}   
\setcounter{equation}{0}   

The Higgs potential in the standard model of particle physics is
\be  
  V_{Higgs} = 
    M_{{\cal H}}^2 {\cal H}^{\dag}{\cal H} + 
    {{\lambda _{{\cal H}}} \over 2} 
      \left( {{\cal H}^{\dag}{\cal H}} \right)^2
\label{higgspotential}
\quad , 
\ee 
where the broken phase, 
corresponding to  
$M_{{\cal H}}^2 < 0$, is chosen in nature. 
This mass parameter and coupling $\lambda _{{\cal H}}$ are 
arranged so that
\be 
  \left\langle {{\cal H}^0} \right\rangle_{broken} \approx 
    { {246 \, GeV} \over {\sqrt{2}} } \approx 175\,GeV
\quad , 
\label{higgsexpectationvalue}
\ee 
in order to produce the experimental values of the masses 
of the $W$ and $Z$. 

In this work, 
we consider the standard model 
in the phase  
for which $M_{{\cal H}}^2 > 0$. 
This is more than an exercise. 
Insight into the physics of the standard model  
can be gained by analyzing it 
in other regions of parameter space. 
Furthermore, 
at high temperatures 
such as those that occurred in the early universe, 
thermal effects added 
a positive contribution to the Higgs mass 
effectively reversing its sign. 
Thus, 
some of the results of our current work 
are relevant to particle physics at high temperatures 
and to early universe cosmology. 
In particular, 
it is a better approximation to use zero for the current algebra masses 
of leptons and quarks 
(including the top quark!) 
at temperatures above $200 \, GeV$ 
than to use the values found in nature at zero temperature. 

We refer to the usual case of $M_{{\cal H}}^2 < 0$ 
as the {\it broken phase}, 
and to $M_{{\cal H}}^2 > 0$ 
as the {\it non-breaking-Higgs-potential phase}. 
Actually, the non-breaking-Higgs-potential phase still 
undergoes electroweak breaking but dynamically. 
The breaking effects are quite small. 
In comparing the two phases, 
we fix a mass scale of QCD rather 
than fixing the strong interaction coupling constant $g_s$. 

Our definition of the standard model  
is the $SU_c (3) \times SU_L (2) \times U_Y (1)$ gauge theory 
with three generations of quarks and leptons. 
It incorporates the Weinberg-Salam-Glashow electroweak model 
including a single Higgs doublet and its Yukawa couplings, 
even though this sector has not been experimentally confirmed. 
If no fundamental Higgs is discovered, 
then it is likely that the true electroweak symmetry breaking mechanism 
will be similar to that created by a Higgs field 
and it is possible that quark and lepton mass generation 
will be mimicked by a phenomenological Higgs field 
and Yukawa interactions. 

It turns out that the ``standard model in its other phase'' 
exhibits some interesting physics. 
Spontaneous dynamical symmetry breaking of the electroweak group 
$SU_L (2) \times U(1)$ takes place 
so that the $W$ and $Z$ still achieve masses, 
although the masses are much smaller 
than in the usual broken phase, 
a result noted by L.\,Susskind 
in his initial paper on technicolor.\ct{susskind} 
Quarks and leptons obtain tiny masses 
by a mechanism identical to that used 
in bosonic technicolor.\ct{bosonictechnicolor} 
In effect, bosonic technicolor is almost realized in nature. 
Had the deconfinement temperature of the strong interactions been 
much higher, 
there would have been a period of the early universe 
dominated by bosonic technicolor! 

We analyze the model in stages: 
We first ignore electroweak effects and Yukawa interactions. 
The electroweak contributions are incorporated next, 
and the Higgs field and its interactions are treated 
as a final perturbation. 

Section II analyzes 
the global symmetries of pure QCD, 
how they are dynamically broken by the strong interactions
and the way in which they are explicitly broken by electroweak effects. 
Dynamical symmetry breaking arises through quark condensates. 
We assume that  
$\left\langle \bar q_i q_i \right\rangle \ne 0$, 
so that axial currents are broken. 
A posteriori, we find this to be consistent: 
Had the wrong pattern of symmetry breaking been assumed, 
certain pseudo-Goldstone bosons 
would have acquired negative square masses.\ct{dashen} 

The spontaneously breaking of axial symmetries 
implies the existence of Goldstone bosons. 
The charged ones acquire small masses 
through electroweak interactions. 
In Section III, 
we compute these one-loop effects. 
One interesting result 
is a cancellation between photon and $Z^0$ contributions, 
so that charged Goldstone boson masses are an order 
of magnitude smaller than one might have expected.  

The neutral Goldstone bosons obtain even smaller masses 
after masses for the quarks and leptons are generated 
by the bosonic-technicolor mechanism: 
When $\left\langle \bar q_i q_i \right\rangle \ne 0$, 
terms linear in the neutral Higgs field are produced, 
and it acquires a vacuum expectation value. 
This value expectation value, in turn, 
leads to quark and lepton masses in the same manner 
as in the Weinberg-Salam-Glashow model. 
However, because these effects are all tiny, 
quark and lepton masses are orders of magnitude 
smaller in the non-breaking-Higgs-potential phase 
than in the breaking phase. 
Section IV calculates the masses. 

In Section V, 
we compute the light hadron spectrum 
drawing on a variety of methods including 
lattice QCD, experimental data and the quark model. 
There is a rich spectrum of lightest baryons 
not only involving up and down quarks 
but the other four quarks too. 
As a result, 
there is an explosion in the number of possible nuclei. 
See Section VI. 
This section also discusses 
the particle physics,
the atomic physics 
and the cosmology 
of the standard model 
in the non-breaking-Higgs-potential phase. 

Here is a partial list of other results: 
At tree level, 
the Weinberg weak-mixing angle 
and the Cabibbo-Kobayashi-Maskawa matrix 
in the non-breaking-Higgs-potential phase
are the same as in the breaking phase. 
The masses of the quarks and leptons 
in the non-breaking-Higgs-potential phase
range from about a milli-electron Volt 
for the electron to several hundred electron Volts 
for the top quark. 
The masses of the lightest vector meson states 
and the spin $3/2$ baryons  
remain about the same. 
There are, however, many more of these states. 
The masses of the lightest baryons are lower by about $3 \%$, 
and the neutron is lighter than the proton. 

\vglue 0.2cm
{\bf\large\noindent II.\ The Pattern of Symmetry Breaking}\vglue 0.2cm
\setcounter{section}{2}   
\setcounter{equation}{0}

This section analyzes the global symmetries of the standard model. 
Some symmetries 
are spontaneously broken dynamically by the strong interactions. 
Since the strong interactions are dominant, 
we shall initially ignore electroweak effects and 
the scalar Higgs field. 
The classification of the global symmetries 
of the quark fields in the 
$SU(3) \times SU(2) \times U(1)$ theory
seems not to have appeared in the literature before, 
although the results of this section 
are a straightforward application of the techniques 
of S.\,Weinberg's paper on dynamical symmetry breaking.\ct{weinberg76}

It is convenient to assemble the Dirac fields 
of the six flavors of quarks into one column vector $\Psi$ as
\be
\Psi =\left( \matrix{u\hfill\cr
  d\hfill\cr
  c\hfill\cr
  s\hfill\cr
  t\hfill\cr
  b\hfill\cr} \right)
\quad . 
\label{Psi}
\ee 
The strong interactions have a global symmetry group  
given by 
$
  S_{strong} = 
   SU_L\left( 6 \right) \times SU_R\left( 6 \right) 
    \times U_V\left( 1 \right)
$,
where the $SU_L\left( 6 \right)$ left-handed currents 
are 
\be
  J_L^{a\mu } = 
   \bar\Psi \gamma^\mu \Lambda^a 
     {{\left( {1-\gamma_5} \right)} \over 2}\Psi 
\quad ,  
\label{lefthandedcurrent}
\ee 
the $SU_R\left( 6 \right)$ currents are 
\be
  J_R^{a\mu } = \bar \Psi \gamma^\mu \Lambda^a 
   {{\left( {1+\gamma_5} \right)} \over 2}\Psi 
\quad , 
\label{righthandedcurrent}
\ee 
and the vector $U_V \left( 1 \right)$ current, 
which is proportional to baryon number, 
is 
\be
  J_V^{U\left( 1 \right)} = 
     3 J_{baryon\;number} = 3 J_b = \bar \Psi \gamma^\mu \Psi 
\quad .
\label{uonevectorcurrent}
\ee
In eqs.\bref{lefthandedcurrent} and \bref{righthandedcurrent}, 
$
  \Lambda^a 
$ 
is an element of the Lie algebra of 
the group $SU\left( 6 \right)$. 
The symmetry associated with the $U_A \left( 1 \right)$ current is absent 
due to the axial anomaly. 
The group $S_{strong}$ is $71$ dimensional. 

It is well known that the strong interactions spontaneously break 
$S_{strong}$ down to a subgroup $H$ 
\be 
   SU_L\left( 6 \right) 
    \times SU_R\left( 6 \right) 
       \times U_V\left( 1 \right) 
   \to SU_V\left( 6 \right)\times U_V\left( 1 \right)
\label{strongbreaking}
\ee
due to the formation 
of quark condensates: 
\be
  \left\langle {\bar \Psi \Psi } \right\rangle \ne 0
\quad , 
\label{condensatebreaking}
\ee
so that the unbroken strong interaction group is  
$
  H = 
SU_V\left( 6 \right)\times U_V\left( 1 \right)
$. 
The axial generators 
\be 
   J_A^{a \mu} = 
  \bar \Psi \gamma^\mu \Lambda^a\gamma_5\Psi 
\label{axialcurrent}
\ee
are spontaneously broken, 
while the vector generators 
\be 
 J_V^{a \mu} =
  \bar \Psi \gamma^\mu \Lambda^a\Psi 
\label{vectorcurrent}
\ee 
and $J_V^{U\left( 1 \right)}$ survive. 
The corresponding broken charges are 
\be
  Q_A^a = 
   \int {d^3x}\bar \Psi \left( x \right)\gamma^0 
         X^a\gamma_5\Psi \left( x \right)
\quad , 
\label{axialcharge}
\ee 
where 
$X^a \in SU \left( 6 \right)$. 
Since there are $35$ generators for $SU \left( 6 \right)$, 
there are $35$ massless Goldstone bosons 
at this stage. 
These states are the analogs of the $0^{-}$ 
scalars of the usual standard model: 
the pions, kaons, $D$'s, $B$'s, $D_s$'s, $B_s$'s, etc., 
as well as bosons involving $t$ quarks 
that do not exist in nature. 

It is convenient to introduce a compact notation 
for the currents. 
They all may be expressed as 
\be
  J_\Sigma^\mu =\bar \Psi \gamma^\mu \Sigma \Psi 
\quad , 
\label{sigmanotation}
\ee
where $\Sigma$ is of the form 
\be
  \Sigma \in U_{U-D} \left( 2 \right) 
     \times U_G\left( 3 \right)\cdot \Gamma 
\quad , 
\label{sigmaform}
\ee
where $U_{U-D}$ is the unitary group 
that acts on up/down type 
quarks, $U_G$ is the group that acts on the three generations, 
and  
$\Gamma$ is a linear combination of $\gamma_5$ and $1_4$ 
($1_4$ is the $4 \times 4$ unit matrix in Dirac space). 
For example, 
the baryon number current $J_b$ is associated with  
\be
  \Sigma_{b} = 
   \;{1 \over 3} I_2 \times I_3 \cdot 1_4
\quad , 
\label{sgimabaryon}
\ee
where 
$I_n$ is the $n \times n$ identity matrix. 

The global symmetry group $S_{strong}$ 
can be partitioned 
into six classes, 
which we label A, B, C, D, E and F 
and which we now define. 

Introduce the $SU_L\left( 2 \right) \times U_Y\left( 1 \right)$ 
electroweak gauge interactions. 
The vector gauge bosons of $SU_L\left( 2 \right)$ 
couple to a current associated with 
\be
  \Sigma_{SU_L\left( 2 \right)} = 
      \;\tau^a\times I_3\cdot {{\left( {1-\gamma_5} \right)} \over 2} 
\quad {\rm (class \ C)} 
\quad , 
\label{sigmac}
\ee
while that of $U_Y\left( 1 \right)$ couples to
\be
  \Sigma_{U_Y\left( 1 \right)} = 
  \left( {\matrix{{1 \over 3}\hfill\cr
  0\hfill\cr}\;\matrix{0\hfill\cr
   {1 \over 3}\hfill\cr}} \right) \times 
     I_3\cdot {{\left( {1-\gamma_5} \right)} \over 2} + 
  \left( {\matrix{{4 \over 3}\hfill\cr
    0\hfill\cr}\;\matrix{0\hfill\cr - 
     {2 \over 3}\hfill\cr}} \right) \times 
       I_3\cdot {{\left( {1+\gamma_5} \right)} \over 2}
\quad . 
\label{sigmauyone}
\ee 
As is well known, 
the electromagnetic current is a linear combination 
of $U_Y\left( 1 \right)$ 
and the ``$\tau^3$'' generator of $SU_L\left( 2 \right)$:   
\be
  \Sigma_{U_{EM}} = 
    \;\left( {\matrix{{2 \over 3}\hfill\cr
     0\hfill\cr}\;\matrix{0\hfill\cr - 
      {1 \over 3}\hfill\cr}} \right)\times I_3\cdot 1_4
\quad . 
\label{sigmaem}
\ee

Some of the symmetries of $S_{strong}$ are also symmetries of 
the electroweak interactions. 
It is straightforward to find those that 
preserve $SU_L\left( 2 \right) \times U_Y\left( 1 \right)$. 
With the exception of the gauge generators, 
these electroweak preserving symmetries 
are associated with the $\Sigma$ that commute
with $\Sigma_{SU_L\left( 2 \right)}$ 
and $\Sigma_{U_Y\left( 1 \right)}$ 
in eqs.\bref{sigmac} and \bref{sigmauyone}. 
The electroweak global symmetry group $S_W$ has $29$ elements: 
$$
  \tau^a \times I_3 \cdot {{\left( {1-\gamma_5} \right)} \over 2} 
\;,\; a=1, \ 2,\ or \ 3 
\ , \quad
  I_2 \times \lambda^\alpha \cdot {{\left( {1-\gamma_5} \right)} \over 2}
\quad , 
$$
\be
  \tau^3\times I_3\cdot {{\left( {1+\gamma_5} \right)} \over 2}
\ , \quad  
  \tau^3 \times \lambda^\alpha \cdot {{\left( {1+\gamma_5} \right)} \over 2}
\ , \quad 
  I_2\times \lambda^\alpha \cdot {{\left( {1+\gamma_5} \right)} \over 2}
\quad ,
\label{sigmasw}
\ee
$$
  I_2\times I_3\cdot 1_4
\quad ,
$$
where $\lambda^\alpha \in SU_G \left( 3 \right)$.
Of these $29$ generators, 
four are associated with the global symmeties $G_W$ 
of the electroweak gauge group
and are given 
in eqs.\bref{sigmac} and \bref{sigmauyone}. 
We have been unable to find 
this list of the global symmetries 
of the weak interactions in the literature, 
so that this result may be of use to theorists. 

There are $42$ generators orthogonal to those 
in eq.\bref{sigmasw} that correspond to currents explicitly 
broken by electroweak gauge interactions:  
$$
  \tau^a\times \lambda^\alpha \cdot {{\left( {1-\gamma_5} \right)} \over 2}
\quad ,
$$
\be
  \tau^1\times \lambda^\alpha \cdot {{\left( {1+\gamma_5} \right)} \over 2}
\ , \quad  
  \tau^2\times \lambda^\alpha \cdot {{\left( {1+\gamma_5} \right)} \over 2}
\quad ,
\label{sigmaewbroken}
\ee
$$
  \tau^1\times I_3\cdot {{\left( {1+\gamma_5} \right)} \over 2}
\ , \quad  
  \tau^2\times I_3\cdot {{\left( {1+\gamma_5} \right)} \over 2}
\quad .
$$
It is convenient to re-organize these currents 
according to parity. 
The even parity vector currents 
\be
  \tau^a\times \lambda^\alpha \cdot 1_4 
\ , \quad  
  \tau^1\times I_3\cdot 1_4 
\ , \quad  
  \tau^2\times I_3\cdot 1_4  
\quad {\rm (class \ F)} 
\label{sigmaf}
\ee 
are not spontaneously broken by the strong interactions, 
while the odd parity axial currents 
\be
  \tau^1 \times \lambda^\alpha \cdot \gamma_5 
\ , \quad   
  \tau^2 \times \lambda^\alpha \cdot \gamma_5 \quad {\rm (class \ E)}
\label{sigmae}
\ee 
are spontaneously broken. 

The $29$ symmetries of $S_W$ can be divided 
into four classes. 
Class A consists of electroweak gauge currents 
that are not spontaneously broken by the strong interactions.
There is only one member and 
it is the electromagnetic current: 
\be 
  \left( {\matrix{{2 \over 3}\hfill\cr
  0\hfill\cr} \;\matrix{0\hfill\cr
  -{1 \over 3}\hfill\cr}} \right) \times 
  I_3\cdot 1_4\,\leftrightarrow U_{EM} \quad {\rm (class \ A)}
\quad . 
\label{sigmaa}
\ee 
There are nine non-gauge currents that are not broken 
by either the strong interactions or the electroweak interactions. 
They are 
\be 
   I_2 \times 
    I_3\cdot 1_4\,\leftrightarrow U_{V}(1) 
\ , \quad 
  I_2 \times 
    \lambda^\alpha \cdot 1_4\,\leftrightarrow SU_G\left( 3 \right)  
 \quad {\rm (class \ B)} 
\quad , 
\label{sigmab}
\ee
corresponding to baryon number and $SU_G ( 3 )$,
the group that acts on the three-dimensional generation space. 
Particle multiplets can thus be classified according 
to charge, baryon number and the representation of $SU_G ( 3 )$. 

Of the remaining $19$ global symmetries 
of the electroweak interactions, 
three correspond to gauge symmetries that are spontaneously 
broken by the strong interactions. 
They constitute Class C and 
are given in eq.\bref{sigmac}. 
The broken currents 
$  
  \tau^a \times I_3\cdot \gamma_5 
$  
can be expressed as a linear combination 
of these three gauge currents 
(eq.\bref{sigmac}) 
and currents conserved 
by the strong interactions. 
In such a case, 
it is well-known\ct{weinberg76} 
that the Goldstone bosons 
corresponding to the broken charges 
$ 
  Q_A^a = 
    \int {d^3x} \bar \Psi \left( x \right)\gamma^0\tau^a \times 
     I_3\gamma_5\Psi \left( x \right) 
$ 
are ``eaten'' by electroweak gauge bosons 
to become massive vector particles 
via the Higgs mechanism. 
Weinberg calls these types of Goldstone bosons 
fictitious because they do not appear 
as scalar particles in the physical spectrum. 
Of course, 
this is dynamical gauge symmetry breaking,\ct{dsb1,dsb2,dsb3,weinberg76} 
which is the basis 
for technicolor.\ct{susskind,weinberg79} 

It turns out that 
the same linear combinations of neutral electroweak gauge bosons 
acquire masses as in the case 
of the usual broken standard model, 
and that the Weinberg mixing angle $\theta_w$ 
is the same: 
$ 
  \sin \left( {\theta_W} \right) = 
   {{g'} \over {\sqrt {g_2^2+g'^2}}}
$.\ct{susskind,weinberg79}  
Thus, the pattern of electroweak breaking is 
identical for non-breaking and breaking Higgs-potential phases. 
The main difference is that 
the masses of the $W$ and $Z$ are smaller,  
the scale being set by the strong interactions: 
\be
  M_W = 
   \sqrt {n_d}  \, {{{g_2} \over 2}} f_\pi \approx 
      {50 \, MeV}  
\ , \quad 
  M_Z = 
  {{M_W} \over {\cos \left( {\theta_W} \right)}} 
   \approx 57\,MeV
\quad , 
\label{intermediatevectormasses}
\ee 
where $n_d = 3$ is the number of electroweak doublet condensates. 
Here, $g_2$ is the $SU_L (2)$ gauge coupling 
(see eq.\bref{ewlagrangian} below) and 
the pion decay constant is normalized as 
$ 
  f_\pi \approx 93\,MeV
$.  
The electroweak $\rho_{EW}$ parameter 
$
{{M_W^2} \over {M_Z^2\cos^2\left( {\theta_W} \right)}}
$ 
is thus $1$ at tree level.\ct{susskind,weinberg79}  
When loop corrections are taken into account, 
$\rho_{EW}$ in the non-breaking-Higgs-potential phase will differ 
slightly from 
$\rho_{EW}$ in the breaking phase. 

We have accounted for all the currents 
except for $16$ 
corresponding to symmetries of $S_W$ 
that are not global versions of electroweak gauge interactions 
and that are broken by the strong interactions. 
They are 
\be 
  I_2 \times \lambda^\alpha \cdot \gamma_5 
\ , \quad  
  \tau^3 \times 
    \lambda^\alpha \cdot {{\left( {1+\gamma_5} \right)} \over 2} 
\quad . 
\label{sigmado}
\ee 
The currents 
\be 
 I_2 \times \lambda^\alpha \cdot \gamma_5 
\ , \quad  
 \tau^3 \times \lambda^\alpha \cdot \gamma_5 
\quad {\rm (class \ D)}
\label{sigmad}
\ee
can be written as a linear combination of 
the currents in eq.\bref{sigmado} 
and currents not broken by the strong interactions. 
It follows that the electrically neutral bosons associated 
with the spontaneously broken charges 
$  
  \int {d^3x} 
   \bar \Psi \left( x \right)\gamma^0I_2 \times 
    \lambda^\alpha \gamma_5\Psi \left( x \right) 
$  
and 
$ 
  \int {d^3x} 
   \bar \Psi \left( x \right)\gamma^0\tau^3 
     \times \lambda^\alpha \cdot \gamma_5\Psi \left( x \right)  
$ 
do not achieve masses through electroweak interactions. 
At this stage, 
they are massless Goldstone bosons, 
or true Goldstone bosons in the language of Weinberg.\ct{weinberg76}

Figure 1 summarizes the partitioning of $S_{strong}$ 
into the six classes. 
Classes A and C are the global versions of the electroweak 
gauge symmetries with class C being spontaneously broken 
by the strong interactions. 
The union of classes A, B, C and D constitutes 
the global symmetry group $S_W$ 
of the electroweak interactions, 
classes B and D being non-electroweak gauge symmetries. 
Class B differs from class D in that it is not spontaneously 
broken by the strong interactions. 
Classes A, B and F compose $H$,
the unbroken strong interaction group. 
Members of F are not symmetries 
of the electroweak gauge interactions.
The complement of $H$, that is, 
the union of classes C, D and E 
are symmetries spontaneously broken by the strong interactions. 
There are potentially Goldstone bosons 
associated with these three classes. 
Class E differs from classes C and D in 
that these symmetries are not respected by 
the electroweak gauge interactions. 
The bosons associated with class C are fictitious in that 
they are eaten by the electroweak vector gauge bosons 
through the Higgs mechanism. 
The bosons of class E are pseudo Goldstone 
because the achieve masses through the electroweak gauge interactions. 
Finally, class D is associated with true Goldstone bosons, 
particles with zero mass when Higgs-Yukawa interactions 
are neglected. 

\vglue 0.2cm
{\bf\large\noindent III.\ The Goldstone Boson Spectrum}\vglue 0.2cm
\setcounter{section}{3}   
\setcounter{equation}{0}   

Of the $35$ Goldstone bosons 
associated with classes C, D and E, 
the three in class C are eaten by the $Z$ and $W$, 
the $16$ electrical neural ones (class D) are massless 
and the $16$ charged ones (class E)
acquire masses from electroweak interactions. 
The purpose of this section is to compute the common 
mass of this charged octet of $SU_G (3)$
using current algebra. 
Although the results of this section 
follow straightforwardly from technicolor methods, 
this is the first time that the methods have 
been applied to the standard model 
in its non-breaking-Higgs-potential phase. 
The formalism for this calculation 
exists in many places. 
We make use of the techniques 
developed by M.\,Peskin and J.\,Preskill 
in refs.\ct{peskin,preskill}. 
We shall review just enough of this formalism 
to set notation and to render the text readable. 

The electroweak gauge bosons interact with 
quarks through the following lagrangian 
\be 
  {\cal L}_{EW} = 
   {{g_2} \over 2}\sum\limits_{a=1}^3  
     A_\mu^a J_{\Sigma_{SU_L\left( 2 \right)}^a}^\mu + 
     {{g'} \over 2} B_\mu J_{\Sigma_{U_Y\left( 1 \right)}}^\mu 
\quad , 
\label{ewlagrangian}
\ee 
where  
$A_\mu^a$ and $B_\mu$ are respectively the gauge bosons 
for the $SU_L (2)$ and $U_Y \left( 1 \right)$ gauge groups, 
which couple to their corresponding currents 
$J_{\Sigma_{SU_L\left( 2 \right)}^a}^\mu$ and 
$J_{\Sigma_{U_Y\left( 1 \right)}}^\mu$ 
(given in eqs.\bref{sigmanotation}, \bref{sigmac} and \bref{sigmauyone}) 
with strengths $g_2 / 2$ and $g' /2$. 
Following ref.\ct{peskin}, 
we rewrite ${\cal L}_{EW}$ 
in terms of left-handed Dirac fields as 
\be 
     {\cal L}_{EW} = \sum\limits_\Xi g_\Xi A_\mu^\Xi J_\Xi^\mu 
\quad ,  
\label{lagrangianform}
\ee 
by using the charge conjugates of right-handed fields. 
Let $\psi$ be a column vector of $12$ entries 
with the first six given by the left-handed components 
of $\Psi$ and the second six given by the conjugates 
of the right-handed components of $\Psi$. 
The vector fields $A_\mu^\Xi$  
in eq.\bref{lagrangianform} 
couple to the currents 
\be 
  J_\Xi^\mu = 
   \bar \psi \gamma^\mu \Xi {{\left( {1-\gamma_5} \right)} \over 2}\psi 
\quad , 
\label{currentform}
\ee 
where $\Xi$ is a matrix in flavor space. 
There are four terms 
in eq.\bref{lagrangianform} 
corresponding to the four vector gauge bosons 
of the electroweak interactions. 
It is convenient to choose these gauge bosons 
to be in a mass diagonal basis, that is, 
to be $\gamma$, $W^\pm$ and $Z^0$. 

The next step in the computation is to write
\be 
  \Xi = \Xi_T+\Xi_X
\quad , 
\label{xidecomposition}
\ee 
where $\Xi_T \in H$ is a generator not broken by the strong interactions, 
and
$ 
  \Xi_X \in S_{strong}/H
$ 
is a broken generator. 

Define the Goldstone decay constant $f_\Pi^a$ by 
\be 
  \left\langle {\Omega |} \right. J_A^{a\mu } {|\Pi^b} \left. \right\rangle = 
   i p^\mu f_\Pi^a \delta^{ab}
\quad , 
\label{fpidefinition}
\ee 
where $\Omega$ is the vacuum state 
and 
$ |\Pi^b \left. \right\rangle $ 
is the Goldstone boson of momentum $p^\mu$ 
associated with the broken charge $Q_A^b$.  
The current algebra $SU (6)$ matrices in $J_A^{a\mu }$ and $Q_A^b$
of eqs.\bref{axialcurrent} and \bref{axialcharge}
are normalized so that
\be 
  Tr\left( {\Lambda^a \Lambda^b} \right) = \delta^{ab}
\ , \quad  
   Tr\left( {X^a X^b} \right) = \delta^{ab} 
\quad . 
\label{matrixnormalization}
\ee

The contribution to the Goldstone boson mass matrix $m_{ab}$ 
from a symmetry breaking perturbation $\delta {\cal H}$ 
is given 
by Dashen's theorem\ct{dashen63} 
\be 
  m_{ab}^2 = 
   -{1 \over {f_\Pi^a f_\Pi^b}} 
    \left\langle {\Omega |} \right. 
     \left[ { Q_A^a , \left[ {Q_A^b , \delta {\cal H} } \right] } \right] 
      \left. {|\Omega } \right\rangle
\quad . 
\label{dashensformula}
\ee 
In the present case, 
$\delta {\cal H}$ arises in second order perturbation theory 
through the one-loop exchange of electroweak vector gauge bosons. 
Figure 2 shows the diagram. 
The shaded region represents all possible QCD interactions. 
These involve sea quark loops and the exchanges of gluons. 
Thus, the computation is non-perturbative in the strong interactions. 

Neglecting electroweak interactions, 
one has
\be 
  f_\Pi^a = f_\Pi 
\quad , 
\label{fpiequality}
\ee 
and eq.\bref{fpiequality} 
is expected to hold quite well even when 
non-strong-interaction perturbations are included. 

Using the above formalism, 
one is able to separate out the group theory factors 
in eq.\bref{dashensformula} via \ct{peskin,preskill}  
\be 
  m_{ab}^2 = 
   {1 \over {4\pi }} \sum\limits_\Xi 
    g_\Xi^2\Delta^\Xi 
     Tr\left( {\left[ {\Xi_T,\left[ {\Xi_T,\Lambda^a} \right]} \right]\Lambda^b - 
      \left[ {\Xi_X,\left[ {\Xi_X,\Lambda^a} \right]} \right]\Lambda^b} \right)
\quad , 
\label{bosonmassseparation}
\ee 
where $\Delta^\Xi$, a parameter of dimension mass squared, 
involves the effects of the propagator 
$\Delta_{\mu \nu }^\Xi \left( x \right)$ of the vector 
boson $A_\mu^\Xi$ in the following time-ordered expectation: 
\be 
  \Delta^\Xi = 
   {{4\pi } \over {f_\Pi^2}}\int {d^4x} 
    \Delta_{\mu \nu }^\Xi \left( x \right) 
    \left\langle {\Omega |} \right.T\left( {J_V^\mu 
     \left( x \right)J_V^\nu \left( 0 \right) - 
      J_A^\mu \left( x \right)J_A^\nu \left( 0 \right)} \right) 
       \left. {|\Omega } \right\rangle
\quad . 
\label{deltaxi}
\ee 
Here, $J_V^\mu$ and $J_A^\mu$ are 
any normalized vector and axial currents. 
For example, 
one can take them to be the ones appearing 
in the third generator of $SU_L (2)$: 
\be 
  J_V^\mu = 
   {1 \over {\sqrt 6}}\bar \Psi \gamma^\mu \tau^3\times I_3\Psi 
\ , \quad 
  J_A^\mu = 
   {1 \over {\sqrt 6}}\bar \Psi \gamma^\mu \tau^3\times I_3\gamma_5\Psi 
\quad . 
\label{samplecurrents}
\ee 

It turns out that, 
of the four contributions 
in the sum over $\Xi$ 
in eq.\bref{bosonmassseparation}, 
the two associated with $W^\pm$ vanish. 
The contribution from the photon 
to a positively charge Goldstone boson $\Pi^+$ is 
\be 
  m_{\Pi^+}^2 = \alpha_{EM} \Delta^\gamma 
\quad . 
\label{emchargemass}
\ee 

It remains to determine the non-perturbative 
parameter $\Delta^\gamma$. 
It  has been accurately estimated analytically 
in ref.\ct{das67},  
but it may be computed using the experimental spectrum 
of the pion: 
\be 
  m_{\pi^+}^2 - m_{\pi^0}^2 = 
   \alpha_{EM}\Delta^\gamma \quad \Rightarrow 
    \quad \Delta^\gamma \approx \left( {415\,MeV} \right)^2
\quad . 
\label{pionmassdifference}
\ee 
{}From eqs.\bref{emchargemass} and \bref{pionmassdifference}, 
one finds that the photon contribution to 
$m_{\Pi^+}$ is approximately $35$ $MeV$. 
This is what one would expect: $m_{\Pi^+}$ should 
be of order of ${\sqrt {\alpha } } \Lambda_{QCD}$, 
where $\Lambda_{QCD}$ is a QCD scale, 
which we take to be $300$ $MeV$. 

However, it turns out that the contribution of $Z^0$ cancels 
most of that of the photon. 
The final result from electroweak interactions is 
\be 
  m_{\Pi^+}^2 = 
  \alpha_{EM}\left( {\Delta^\gamma - \Delta^Z} \right)
\quad . 
\label{ewchargemass}
\ee 
Since the mass of $Z^0$ is small compared to $\Lambda_{QCD}$, 
$\Delta^\gamma$ and $\Delta^Z$ 
of eq.\bref{ewchargemass} 
are almost equal. 
In fact, the difference vanishes as $M_{Z}^2$. 
Thus, $m_{\Pi^+}$ should be of order 
${\sqrt {\alpha } } M_{Z}$, or about $5$ $MeV$,  
an order of magnitude smaller than without the cancellation. 

This cancellation should not be surprising. 
A similar effect\ct{peskin,preskill} 
occurs in technicolor models 
for the pseudo Goldstone boson often refered to 
as $P$.\ct{eichtenlane} 
Unfortunately, 
there is no known non-perturbative way 
to compute $\Delta^Z$. 
However, 
eq.\bref{ewchargemass} can be calculated in perturbation theory. 
One finds
\be 
  m_{\Pi^+}^2 \approx 
  {{3\alpha_{EM}} \over {4\pi }} M_Z^2 
    \ln \left( {{{\Lambda_{QCD}^2} \over {M_Z^2}}} \right) \sim 
     \left( {4\,MeV} \right)^2
\quad .
\label{ewchargemasspt}
\ee

In summary, 
the eight Class-E charged pseudo Goldstone bosons, 
the analog of the charged pion, $K$, $D$, $B$, $D_s$, 
etc.\ of the standard model, 
obtain masses of about $4$ $MeV$ 
through the electroweak interactions.

\vglue 0.2cm
{\bf\large\noindent IV.\ The Effects of the Higgs Sector}\vglue 0.2cm
\setcounter{section}{4}   
\setcounter{equation}{0}

No bare quark masses have been generated 
by the electroweak gauge interactions, 
so that quarks are massless at this stage. 
When the Higgs sector along with its Yukawa 
couplings are included, 
then tiny quark and lepton masses arise 
through a mechanism identical to that  
of bosonic technicolor.\ct{bosonictechnicolor}
The purpose of this section 
is to determine the quark and lepton mass spectrum 
for the standard model in its other phase.

The coupling of the $SU_L (2)$ Higgs double 
$ 
  {\cal H} = 
    \left( \matrix{{\cal H}^+ \hfill\cr
     {\cal H}^0\hfill\cr} \right)
$ 
to quarks and leptons in the standard model is
\be 
{\cal L}_{Yukawa} = 
  \sum\limits_{i,j=1}^3  
   \left( {\lambda _{ij}^D\bar Q_L^i{\cal H} D_R^j + 
    \lambda _{ij}^U\bar Q_L^i \tilde {\cal H} U_R^j + 
     \lambda _{ij}^E\bar L_L^i{\cal H}E_R^j} \right) + {\rm {c.c.}}
\quad , 
\label{yukawalagrangian}
\ee 
where $i$ and $j$ denote different generations; 
$\lambda _{ij}^D$, $\lambda _{ij}^U$ and $\lambda _{ij}^E$ 
are Yukawa coupling constants; 
$Q_L^i$ and $L_L^i$ are the left-handed $SU_L (2)$ doublets 
of quarks and leptons for the $i$th generation; 
$D_R^j$, $U_R^j$ and $E_R^j$ are right-handed $SU_L (2)$ singlets
of down quarks, up quarks and charged leptons for the $j$th generation; 
and  
$ 
  \tilde{\cal H} = 
    \left( \matrix{{\cal H}^{0*}\hfill\cr
      -{\cal H}^-\hfill\cr} \right)
$ 
is the $SU_L (2)$ conjugate Higgs field.  

When $\bar q q$ quark condenstates form, 
the Yukawa interactions generate linear terms for ${\cal H}$, 
thereby causing the Higgs field to acquire a vacuum expectation value: 
The neutral component of ${\cal H}$ must be shifted 
to eliminate the linear terms.  

Although the bosons absorbed by the $Z^0$ and $W^\pm$ 
now involve a tiny add-mixture of Higgs, 
the Higgs survives, 
and there are two neutral bosons 
and one charged boson of mass of the order of $M_{{\cal H}}$.

When the vacuum expectation of ${\cal H}^0$ is substituted 
into ${\cal L}_{Yukawa}$, 
the quarks and charged leptons obtain masses.  
The same unitary transformations on matter fields 
as in the broken phase 
diagonalize mass matrices. 
It follows that, at tree-level, 
the Cabibbo-Kobayashi-Maskawa matrix 
in the non-breaking-Higgs-potential phase 
is the same as in the broken phase. 
The mixing-angle effects enter the charged gauge weak interactions 
but do not affect the pseudo Goldstone boson computations 
in Section III, 
because the $W^\pm$ do not contribute,
and even if they did, 
the mixing angles are small. 

The calculation of $\left\langle {{\cal H}^0} \right\rangle$ 
is straightfoward.
The shift is quite small 
so that the quartic term in $V_{Higgs}$ may be neglected.
One finds 
\be 
  M_{{\cal H}}^2 \left\langle {{\cal H}^0} \right\rangle \approx 
   \sum\limits_{k=1}^6 \lambda_{q_i} \left\langle {\bar q_iq_i} \right\rangle
\quad , 
\label{higgsshift}
\ee 
where $\lambda_{q_i}$ are the diagonalized Yukawa couplings 
for quarks 
($m_{q_i} = \lambda_{q_i} \left\langle {{\cal H}^0} \right\rangle$).
Although $\left\langle {\bar q_iq_i} \right\rangle$ is known, 
$ 
  \left\langle {\bar q_i q_i} \right\rangle \approx 
    \left( {225\,MeV} \right)^3
$,
the mass of the Higgs field is not. 
Therefore, 
only an order of magnitude estimate 
for $\left\langle {{\cal H}^0} \right\rangle$ 
and quark and lepton masses can be made. 
Assuming 
$ 
  100\,GeV\le M_{{\cal H}} \le 300 \, GeV
$, 
one finds 
\be 
  \left\langle {{\cal H}^0} \right\rangle \sim 100 \,eV \, to \, 1 \,KeV
\quad . 
\label{higgsexpectationestimate}
\ee 
If follows that 
\be 
  {{\left\langle {{\cal H}^0} \right\rangle} \over 
   {\left\langle {{\cal H}^0} \right\rangle_{broken}}} \sim 10^{-9}
\quad , 
\label{higgsexpectationratio}
\ee 
so that quark and lepton masses are about $10^9$ times smaller 
in the non-breaking-Higgs-potential phase 
as in the broken phase. 

For example, 
taking $M_{{\cal H}} = 175 \, GeV$, 
which gives 
$
  \left\langle {{\cal H}^0} \right\rangle / 
   \left\langle {{\cal H}^0} \right\rangle_{broken} \sim 
     2 \times 10^{-9}
$,
one obtains  
$$ 
  m_t \sim 400 \, eV
\ , \quad 
  m_c \sim 3 \, eV
\ , \quad
  m_u \sim 0.01 \, eV
\quad , 
$$ 
\be 
  m_b \sim 10 \, eV
\ , \quad 
  m_s \sim 0.4 \, eV
\ , \quad
  m_d \sim 0.02 \, eV
\quad , 
\label{quarkandleptonmasses}
\ee 
$$ 
  m_\tau \sim 4 \, eV
\ , \quad 
  m_\mu \sim 0.25 \, eV
\ , \quad 
  m_e \sim 0.001 \, eV
\quad . 
$$ 

The spectrum of the neutral pseudo Goldstone bosons of Class E  
is the order of 
$
\sqrt{  \left\langle {{\cal H}^0} \right\rangle / 
   \left\langle {{\cal H}^0} \right\rangle_{broken}  } 
$
($\sim 5 \times 10^{-5}$) 
times the spectrum of the unbroken model. 
With $M_{{\cal H}} = 175 \, GeV$, 
one finds 
$$ 
 {\rm Those \ bosons \ with \ a \ t \ quark:} \ \sim 0.7 \,MeV  
    \quad (5 \ {\rm states})
\quad , 
$$ 
$$ 
 {\rm Those \ bosons \ with \ a \ c \ quark \ but \ no \ t \ quark:} \ \sim 0.07 \,MeV 
    \quad (3 \ {\rm states}) 
\quad , 
$$ 
\be 
 {\rm Those \ bosons \ with \ a \ b \ quark:} \ \sim 0.14 \,MeV 
    \quad (5 \ {\rm states})
\quad , 
\label{neutralbosonmass}
\ee 
$$  
 {\rm Those \ bosons \ with \ an \ s \ quark \ but \ no \ b \ quark:} \ \sim 0.025 \,MeV 
    \quad (3 \ {\rm states})
\quad . 
$$ 
Thus, the masses of the analogs of the neutral $0^-$ states 
of the usual standard model are as follows: 
That of $D^0$ is about $0.07 \, MeV$, 
those of the $B^0$ and $B_s^0$ are about $0.14 \, MeV$, 
and that of the $K^0$ is about $0.025 \, MeV$. 
There is no neutral state ($\pi^0$) involving only u and d quarks:  
Most of it is eaten by the $Z^0$ 
and part of it appears in the other neutral states.

\vglue 0.2cm
{\bf\large\noindent V.\ The Rest of the Light Hadron Spectrum}\vglue 0.2cm
\setcounter{section}{5}   
\setcounter{equation}{0}

In this section, 
we determine the spectrum 
of the lighter non-Goldstone-boson hadrons 
using a combination of methods that include 
the quark model, lattice QCD and experimental data. 
Interesting differences arise between the hadron spectrum
of the standard model 
in its broken and non-breaking-Higgs-potential phases. 

There is one remaining pseudo scalar 
corresponding to the would-be Goldstone boson 
associated with $\bar \Psi \gamma^\mu \gamma_5\Psi $. 
It receives its mass $m_A$ 
from topological fluctuations, 
which can be computed using the Witten formula:\ct{witten}
\be 
  m_A^2 = 
   2L{{\left\langle {Q_{top}^2} \right\rangle} \over {f_\pi ^2}}
\quad , 
\label{axialbosonmass}
\ee
where $L$ is the number of light quarks 
and $\left\langle {Q_{top}^2} \right\rangle$ 
is the $SU_c (3)$ topological susceptibility, 
which can be numerically determined 
from experimental data as  
$
  \left\langle {Q_{top}^2} \right\rangle 
    \approx \left( {180\;MeV} \right)^4
$. 
There is also a contribution to $m_A^2$ 
from topological fluctuations in the $SU_L (2)$ sector 
but it is considerably smaller than the $SU_c (3)$ term. 
Using eq.\bref{axialbosonmass}, 
one finds 
\be 
  m_A \approx 1210\;MeV
\quad . 
\label{axialbosonmassvalue}
\ee 
This mass is considerably larger than that of the $\eta^\prime$ 
in the broken phase 
because $L$ is $6$ rather than $2$ to $3$. 

The hadron spectrum has an approximate $SU_f (6)$ flavor symmetry
due to the $6$ quarks with masses
much less than the scale of QCD. 
The vector mesons consist of $36$ states 
transforming as an adjoint ${\ul {35}}$ and singlet of $SU_f (6)$. 
They all have approximately the same mass $m_V$, 
which can be determined as follows. 
In the quark model, 
the splitting between scalar and vector states 
is due to the color hyperfine interaction 
between pairs of quarks: 
\be
 \Delta H_{color\;hyperfine} = 
    -{{2\pi \alpha _s} \over {3m_1m_2}} 
    \delta ^3\left( r \right)\lambda _1\cdot \lambda _2S_1\cdot S_2
\quad , 
\label{colorhyperfine}
\ee 
where $m_i$, $q_i$ and $S_i = \sigma_i / 2$ 
are the constituent mass, charge and spin of the $i$th quark. 
A spin independent meson mass $M_0$ can 
be determined by removing these spin-spin 
interaction effects: 
\be 
  M_0 = {3 \over 4} m_V + {1 \over 4} m_S 
\quad , 
\label{spinindependentmass}
\ee 
where $m_S$ is the mass of the pseudo scalar mesons. 
Using experimental data, 
one finds $M_0 \approx 610 \, MeV$. 
In the non-breaking-Higgs-potential phase, 
the quarks are lighter 
and $M_0 $ should be smaller by about $15 \, MeV$. 
On the other hand, 
the color hyperfine interaction is slightly enhanced 
since the masses $m_1$ and $m_2$ 
in eq.\bref{colorhyperfine} 
are smaller by about $10 \, MeV$. 
Taking these two effects into account, 
we find, for the vector meson mass $m_V$,  
\be 
  m_V \approx 790\;MeV
\quad , 
\label{bosonvectormass}
\ee 
a value slightly larger than the mass 
of the $\rho$ and $\omega$. 
The use of the non-relativistic quark model for the computation 
of the vector mesons is approximate but is reliable. 
See for example Table 1.1 of ref.\ct{samuelmoriarty}.

The ligher baryons consist of a spin-$3/2$ ${\ul {56}}$ of $SU_f (6)$ 
containing the $\Delta$ and its partners, 
and a spin-$1/2$ ${\ul {70}}$  of $SU_f (6)$ 
containing the nucleon and its partners. 
These two states are again split by the hyperfine interaction 
in eq.\bref{colorhyperfine}. 
Using similar methods as in the vector meson case, 
we find 
\be 
  M_{J=3/2} \approx 1217\;MeV
\ , \quad  
  M_{J=1/2} \approx 909\;MeV
\quad . 
\label{baryonmass}
\ee 
The mass of the $J=3/2$ baryons is slightly less than 
the experimental mass of the $\Delta$,  
while the mass of the $J=1/2$ baryons is somewhat less than 
the mass of the proton found in nature. 

Of particular interest 
is the mass of the lightest baryon  
since it will be stable. 
Small mass differences among the $J=1/2$ baryons 
are generated by the electroweak interactions. 
These interactions have an $SU_G (3)$ symmetry. 
Under $SU_G (3)$, 
the ${\ul {70}}$ of $SU_f (6)$ decomposes into 
an octet with electric charge $+2$; 
a singlet, two octets and a decaplet with charge $+1$; 
a singlet, two octets and a decaplet with charge $0$; 
and an octet with electric charge $-1$.
The two $+1$ charged octets 
are distinguished by their properties 
under interchange of the two up-type quarks: 
one is symmetric while the other is antisymmetric. 
Likewise, the two neutral octets 
are distinguished by their properties 
under the interchange of the two down-type quarks. 

In the quark model in its broken phase, 
there are three contributions to baryonic mass differences: 
(a) the up/down quark mass difference, 
(b) the electrostatic potential energy, 
which may be estimated from the pairwise 
interaction of quarks via
\be 
  \Delta H_{static \ potential} = 
   q_1 q_2 \left\langle {{{1} \over {r_{12}}}} \right\rangle
\quad , 
\label{electrostaticenergy}
\ee 
where $q_i$ is the charge on the $i$th quark 
and where 
$\left\langle {{{1} \over {r_{12}}}} \right\rangle$ 
is the average inverse distance between 
the two quarks, 
and (c) the electric hyperfine interaction of 
\be 
  \Delta H_{hf} = 
   -{{q_1 q_2} \over {3m_1 m_2}} 
    \delta ^3\left( r_{12} \right) S_1 \cdot S_2
\quad . 
\label{electrichyperfine}
\ee 
The baryonic mass differences can be fairly reliably 
computed in the quark model. 
For example, 
ref.\ct{samuelmoriarty} 
obtained the above three contributions 
using lattice QCD to compute 
$\left\langle {{{1} \over {r_{12}}}} \right\rangle$ 
and $\psi \left(  {r_{12}=0} \right) $, 
which enters in 
eq.\bref{electrichyperfine} 
when baryonic wave functions $\psi$ are used. 
Reference \ct{samuelmoriarty} 
finds the following results 
\begin{center} 
$$ 
\matrix{{Particles}
  \hfill &{Static\,Term} 
    \hfill &{u-d\,Mass\,Term}
      \hfill &{Total}
       \hfill &{Experiment}\hfill \cr
  {\quad p-n}
    \hfill &{\qquad 0.51} 
     \hfill &{\qquad -2.21}
      \hfill &{-1.69}
       \hfill &{\qquad -1.29}\hfill \cr
  {\;\;\Sigma^+-\Sigma^0} 
    \hfill &{\qquad 0.07}
     \hfill &{\qquad -2.89}
      \hfill &{-2.82}
       \hfill &{\qquad -3.23}\hfill \cr
  {\;\;\Sigma^0-\Sigma^-}
    \hfill &{\quad \;-1.62}
     \hfill &{\qquad -2.89}
      \hfill &{-4.51}
       \hfill &{\qquad -4.81}\hfill \cr
  {\;\;\Xi^0-\Xi^-}
    \hfill &{\quad \;-1.76}
     \hfill &{\qquad -3.47}
      \hfill &{-5.23}
       \hfill &{\quad -6.4\pm 0.6}\hfill \cr
  {\;\;\Sigma^{*+}-\Sigma^{*0}}
    \hfill &{\qquad 0.41}
     \hfill &{\qquad -2.05}
      \hfill &{-1.64}
       \hfill &{\quad -0.9\pm 1.1}\hfill \cr
  {\;\;\Sigma^{*+}-\Sigma^{*-}}
    \hfill &{\quad \;-0.86}
     \hfill &{\qquad -4.11}
      \hfill &{-4.97}
       \hfill &{\quad -4.4\pm 0.7}\hfill \cr
  {\;\;\Xi^{*0}-\Xi^{*-}}
    \hfill &{\quad \;-1.29}
     \hfill &{\qquad -2.32}
      \hfill &{-3.60}
       \hfill &{\quad -3.2\pm 0.7}\hfill \cr}
$$ 
Table 1 Baryonic Mass Splittings for the Broken Phase (in $MeV$)\\ 
\ 
\end{center}
It should be noted that the above calculation 
involved no adjustable parameters 
because input parameters had been fixed in an earlier 
computation of the meson spectrum. 
By comparing theory with experiment, 
Table 1 provides an estimate of the accuracy of our methods. 
In particular, 
the sign of all mass differences is correctly reproduced. 

The baryonic mass splittings for 
the non-breaking-Higgs-potential phase 
proceed as in the broken phase 
with the following differences: 
there is no significant contribution from item (a) above 
because up-type and down-type quarks 
have almost identical masses,   
and there are contributions 
from the $Z^0$ and $W^\pm$. 
Because the baryons involve 
equal admixtures of left and right quarks, 
formulas \bref{electrostaticenergy} and \bref{electrichyperfine} 
hold for $Z^0$ exchange 
with 
$ 
  q \to \left( {q_L + q_R} \right)/2
$. 
The $W^\pm$ only contributes to baryons 
with both up-type and down-type quarks 
and interchanges the two types, 
which must be taken into account. 
We have computed the effects of the exchange of the $Z^0$ and $W^\pm$ 
using the baryonic wave functions 
of ref.\ct{samuelmoriarty}  
and incorporating a suppression factor 
of $exp ( - M_V / \Lambda_{QCD} ) \approx 0.83$ 
due to the non-zero value of the mass $M_V$ 
of these vector gauge bosons. 

Here are the tabulated contributions to the 
$J=1/2$ baryonic mass splittings. 
\begin{center} 
$$ 
\matrix{
 {State}
  \hfill &{\;EM_{sp}}
   \hfill &{\;EM_{hf}}
    \hfill &{{\;\;\;Z_{sp}}}
     \hfill &{\;\;\;Z_{hf}}
      \hfill &{\;\;W_{sp}}
       \hfill &{\;\;W_{hf}} 
        \hfill &{Total}\hfill \cr
  {uuu\ 8}
   \hfill &{\;\;2.4}
    \hfill  &{\;\;0.4}
     \hfill  &{\;\;0.24}
      \hfill  &{\;\;0.04}
       \hfill  &{\;\;\;\;0}
        \hfill  &{\;\;\;\;0}
         \hfill  &{\;\;3.07}\hfill \cr
  {uud\ 10}
   \hfill  &{\;\;\;0}
    \hfill  &{-0.4}
     \hfill  &{-0.20}
      \hfill  &{-0.11}
       \hfill  &{-0.81}
        \hfill  &{-0.27}
         \hfill  &{-1.79}\hfill \cr
  {uud\ 8_+}
   \hfill  &{\;\;\;0}
    \hfill  &{-0.4}
     \hfill  &{-0.20}
      \hfill  &{-0.11}
       \hfill  &{\;\;0.40}
        \hfill  &{\;\;0.13}
         \hfill  &{-0.17}\hfill \cr
  {uud\ 8_-}
   \hfill  &{\;\;\;0}
    \hfill  &{\;\;0.4}
     \hfill  &{-0.20}
      \hfill  &{\;\;0.04}
       \hfill  &{\;\;0.40}
        \hfill  &{\;\;\;\;0}
         \hfill  &{\;\;0.64}\hfill \cr
  {uud\ 1}
   \hfill  &{\;\;\;0}
    \hfill  &{\;\;0.4}
     \hfill  &{-0.20}
      \hfill  &{\;\;0.04}
       \hfill  &{-0.81}
        \hfill  &{\;\;\;\;0}
         \hfill  &{-0.57}\hfill \cr
  {ddu\ 1}
   \hfill  &{-0.6}
    \hfill  &{\;\;0.1}
     \hfill  &{-0.03}
      \hfill  &{\;\;0.13}
       \hfill  &{-0.81}
        \hfill  &{\;\;\;\;0}
         \hfill  &{-1.20}\hfill \cr
  {ddu\ 8_+}
   \hfill  &{-0.6}
    \hfill  &{\;\;0.1}
     \hfill  &{-0.03}
      \hfill  &{\;\;0.13}
       \hfill  &{\;\;0.40}
        \hfill  &{\;\;\;\;0}
         \hfill  &{\;\;0.00}\hfill \cr
  {ddu\ 8_-}
   \hfill  &{-0.6}
    \hfill  &{-0.3}
     \hfill  &{-0.03}
      \hfill  &{-0.14}
       \hfill  &{\;\;0.40}
        \hfill  &{\;\;0.13}
         \hfill  &{-0.53}\hfill \cr
  {ddu\ 10}
   \hfill  &{-0.6}
    \hfill  &{-0.3}
     \hfill  &{-0.03}
      \hfill  &{-0.14}
       \hfill  &{-0.81}
        \hfill  &{-0.27}
         \hfill  &{-2.14}\hfill \cr
  {ddd\ 8}
   \hfill  &{\;\;0.6}
    \hfill  &{\;\;0.1}
     \hfill  &{\;\;0.76}
      \hfill  &{\;\;0.13}
       \hfill  &{\;\;\;\;0}
        \hfill  &{\;\;\;\;0}
         \hfill  &{\;\;1.58}\hfill \cr}
$$
Table 2 Electroweak Contributions to Baryonic Masses (in $MeV$)\\ 
\ 
\end{center}
In this table, $u$ stands for any of the up-type quarks ($u$, $c$ or $t$) 
and $d$ stands for any of the down-type quarks ($d$, $s$ or $b$). 
The octet containing the neutron 
is the lightest state 
weighing about one-third of an $MeV$ 
less than the octet containing the proton. 
One sees from Table 2 that this is  due 
to the difference in electromagnetic energy.  
This result is important for the discussion 
of the nuclear and atomic physics of the 
non-breaking-Higgs-potential phase.

\vglue 0.2cm
{\bf\large\noindent VI.\ The Physics of the Standard Model 
in Its Other Phase}\vglue 0.2cm
\setcounter{section}{6}   
\setcounter{equation}{0}

\noindent{\bf A. Nuclear Physics: A Rich World with Lots of Isotopes}  

We use the term {\it nucleon } to refer to any of the light 
three-quark states including those that involve the s, c, b and t quarks.
In the non-breaking-Higgs-potential phase 
these nucleons can have four charges: $+2$, $+1$, $0$ and $-1$. 
The $+2$ and $-1$ octets probably do not form small nuclear 
bound states because they are heavier than the lightest nucleon 
by more than $3.5 \, MeV$. 
Of the $+1$-charged and neutral nucleons, 
the decaplet has the lowest mass. 
Throughout this section, 
we call $+1$-charged and neutral nucleons 
{\it protons} and {\it neutrons} even though there 
are now many types of each, some of which are comprised 
of strange, charm, bottom and top quarks. 

Compared to the usual standard model, 
the non-breaking-Higgs-potential phase 
has four main differences 
as far as nuclear physics is concerned: 
(i) the neutrons are stable 
and the protons are slightly heavier than the neutrons, 
(ii) the Yukawa forces between nucleons are modified 
because the pseudo-Goldstone bosons are lighter, 
(iii) there are a decaplet of lightest protons $p_{10}$ 
and a decaplet of lightest neutrons $n_{10}$ 
(which we denote simply by $p$ and $n$),
and 
(iv) besides the lightest protons 
and neutrons, there are other nucleons, 
the least massive of which is the $SU_G (3)$ singlet $n_{1}$. 

Concerning (ii), 
one can estimate how the Yukawa forces are changed. 
Single exchange of pseudo-Goldstone bosons  
is suppressed because the amplitude vanishes 
as their mass goes to zero. 
This leads to a weakening of the spin-spin interaction. 
Multiple exchanges of the light bosons leads to stronger 
and longer-ranged forces. 
Thus, the scalar and tensor parts of the two-body nuclear potential
are strengthened. 
Other nuclear potentials are expected to remain the same
or be strengthened.
For example, 
consider the short-ranged repulsive force. 
When nucleons are made from only two types of quarks, 
core repulsion is generated by QCD effects arising from 
spin-spin quark interactions and the Pauli exculsion principle. 
However, with six light quarks, 
certain nucleons can overlap. 
For example, 
an $ssc$ neutron experiences little short-ranged repulsion 
with a $uud$ proton.
In summary, 
one expects nucleons to join to form nuclei with at least  
the same binding energies as in the standard model in its usual phase. 

The above four differences lead to 
an extraodinarily rich spectrum of nuclei. 
In what follows, 
we mostly consider nuclei made from the two decaplets. 

Using charge to distinguish nucleonic elements, 
there are an infinite number of elements, 
each of which has an infinite number of isotopes! 
The lightest $Q=0$ nucleus 
is the spin-$1/2$ decaplet of neutrons. 
Two neutrons also form a bound state. 
There are $45$ spin-$1$ states 
and $55$ spin-$0$ states.  
These can be decomposed into $SU_G (3)$ multiplets: 
$45 = { \ul { \bar {10} } } + {\ul {35} }$; 
$55 = {     \ul {27} } + {\ul {28} }$. 
In nature, 
the spin-spin nuclear interaction favors the formation 
of a spin-$1$ state over a spin-$0$ state. 
However, 
in the standard model in its non-breaking-Higgs-potential phase, 
this spin-spin term is significantly weakened 
so that both spin states are stable. 
The spin-$1$ ${ \ul { \bar {10} } }$ 
has the least amount of core repulsion 
and is the lightest of these di-neutron nuclei. 
Of the $120$ spin-$3/2$ and $330$ spin-$1/2$ tri-neutron nuclei, 
the spin-$3/2$ $SU_G(3)$ singlet is lightest. 

It is clear that any number of neutrons 
can join to form a nucleus.
Thus, there are an infinite number of $Q=0$ isotopes. 
The same is true for $Q>0$ nuclei: 
any number of neutrons can be added to a nucleus 
to produce isotopes. 
It is therefore possible to form giant nucleonic nuggets 
constisting of countless neutrons. 
Roughly speaking, 
they would be similar to neutron stars 
but would have a greater range of sizes, 
anything varying from the microscopic to the macroscopic: 
nuclei of hundreds of neutrons, neutron nuggets, neutron balls 
and neutron planets. 
In large nuclei, 
the binding energy per nucleon 
should be about the same as in the usual standard model: 
around $8 \, MeV$. 

The $Q=1$ protons are not stable; 
they decay into a neutron, a neutrino 
and a positively charged lepton. 
Their lifetimes can be estimated 
to be about $10^{-7}$ seconds 
by using the well-known formulas  
and taking into account the differences in phase space 
and in the strength of the weak interactions. 
However, protons can ``live longer'' in nuclei 
because the proton-neutron mass difference is 
less than the binding energy generated by adding 
a proton to a nucleus. 
Thus, the ``deuterons'' 
consisting of one proton and one neutron exist. 
There are $100$ spin-$1$ states, 
which decompose into irreducible multiples of 
${ \ul { \bar {10} } } + {\ul {27} } + {\ul {28} } + {\ul {35} } $
under $SU_G (3)$. 
The ${ \ul {\bar {10} } }$ is expected to be the lightest. 
The spin-$0$ state, which does not quite exist in nature, 
now arises because 
the repulsive nuclear spin-spin interaction is significantly weakened. 
There are $100$ such spin-$0$ states.  
Among the ``tritium'' $n$-$n$-$p$ states, 
there are  $450$ spin-$3/2$ nuclei 
and $1000$ spin-$1/2$ nuclei.  

Two protons and one neutron bind to form ``$^3$He''. 
The number of such nuclei is 
identical to the ``tritium'' case: 
$450$ spin-$3/2$ states and  
$1000$ spin-$1/2$ states. 
It is evident that, 
in the standard model in its non-breaking-Higgs-potential phase, 
there is a plethora of nuclei. 
For ``$^4$He'', 
there are $5050$ spin-$0$ states, 
$6975$ spin-$1$ states and 
$2025$ spin-$2$ states!
Since enough neutrons can always be added to spread out the charge 
from the protons, 
nuclei of all atomic number $Z$ exist. 

The question arises as to whether charged nuclei are stable or decay. 
Since protons can be transformed into neutrons of different type 
(perhaps through a Cabbibo suppressed process) via the 
weak interactions until smaller charged 
nuclei are obtained and since such nuclei are expected to be lighter, 
it would appear that $Q>0$ nuclei are unstable. 
Additional research is needed to determine the lifetimes 
of these charged nuclei.  

In the usual standard model, 
the $^4$He is particularly stable 
because all four nucleons can be placed 
in the lowest level of the shell model. 
In the non-breaking-Higgs-potential-phase case, 
the analog of $^4$He is $^{40}$Ca$^{20}$. 
Each of the ten spin-up protons, ten spin-up neutrons, 
ten spin-down protons and ten spin-down neutrons 
can be put in the $1S$ level without violating the 
Pauli exclusion principle. 
This nucleus is an $SU_G(3)$ singlet and has spin $0$. 
If the $1S$ and $1P$ levels are maximally filled, 
one obtains the nucleus $^{160}$Hg$^{80}$. 

The singlet $n_1$ 
whose mass is about $0.94 \, MeV$ heavier than 
the decaplet neutron can also participate in nuclear bound states. 
In isolation, 
it decays to a charged lepton, an antineutrino and a proton. 
However, it can also be added to nuclei to form isotopes. 
For example, $n_1$ and $p_{10}$ bind 
to create $10$ heavy deuteron states. 
If the usual deuteron is considered to be ``heavy hydrogen,'' 
then $n_1$-$p_{10}$ would be heavy ``heavy hydrogen.'' 
The spin-up $n_1$ and spin-down  $n_1$ 
can be placed in the same shell leading to new magic number nuclei 
such as $^{42}$Ca$^{20}$, $^{162}$Hg$^{80}$ and $^{168}$Hg$^{80}$.

\noindent{\bf B. Stronger Weak Interactions} 

In terms of particle physics, 
the most pronounced difference%
{\footnote {There are many smaller differences. 
For example, the strong interaction coupling constant 
runs more slowly 
in the non-breaking-Higgs-potential phase 
because there are more light quarks. 
Also, $CP$ violation is likely to be less in this phase 
because the magnitude of the entries in 
the qaurk mass matrix are smaller, 
thereby implying that the phases in 
the Cabibbo-Kobayashi-Maskawa matrix 
have less effect.} 
between the breaking   
and the non-breaking Higgs-potential phases  
is the strength of the weak interactions: 
$G_F$ is about $2.5 \times 10^6$ times bigger 
due to the smaller value of the $W$ mass. 
This leads to an enhancement of about 
$6.7 \times 10^{12}$ in matrix elements squared. 
This is the reason why the protons are so short-lived, 
decaying in about $10^{-7}$ seconds.
On the other hand, 
the phase space for many weak processes is greatly 
reduced or smaller mass parameters enter in decay rates;  
This is actually the dominant effective 
for many light states. 
For example, 
the muon can decay into an electron, 
a muon neutrino and an anti-electron neutrino. 
However, the muon's lifetime is about $2 \times 10^{17}$ years! 
The tau is also long-lived, 
lasting $1.5 \times 10^{11}$ years
and decaying into a muon, a tau neutrino 
and an anti-muon neutrino. 
For the charged pseudo scalar bosons, 
the above two effects roughly cancel. 
The lifetime of these particles 
is about $5 \times 10^{-8}$ seconds. 
They decay into a tau and a tau neutrino. 
The above results are obtained using textbook formulas 
and incorporating the different values of masses and $G_F$.

Because the weak interactions are stronger, 
parity-violating effects will be more pronounced 
in the non-breaking-Higgs-potential phase. 
Such effects show up 
as small admixtures of opposite parity 
in nuclei and atoms. 

\noindent{\bf C. Atomic Physics: Giant Atoms} 

Because of the electron's low mass of a milli-electron Volt, 
it would be very difficult for charged nuclei to capture 
electrons to form atoms. 
If this were to happen, 
the binding energy would be about $10^{-8} \, eV$ 
and the atom would be a dozen centimeters large. 

However, since tau's and muons are long-lived, 
they can bind to nuclei. 
Atoms constructed with tau's replacing electrons 
would be about a dozen microns big 
and have binding energies of about a milli-electron Volt. 
Atoms constructed with muons would be about ten times larger 
and have binding energies of about $10^{-4} \, eV$. 
If stable charged nuclei exist, 
then it would take a long time 
for ``a universe in the non-breaking-Higgs-potential phase''
to sufficiently cool to allow these types of atoms. 

\noindent{\bf D. Cosmology: Plasma Domination, Exotic Leptons 
and Anti-Leptons}

There are a lot of interesting details concerning 
the cosmology of the standard model in its other phase. 
However, in this subsection, 
we focus on some general features. 

Up to a trillionth of a second, 
there are no essential changes 
because high temperatures prevent electroweak breaking. 
In the non-breaking-Higgs-potential phase, 
the $SU_L (2) \times U_Y (1)$ breaking 
occurs when confinement sets in 
at about a millionth of a second 
when the temperature $T$ of the universe is around $200 \, MeV$. 
Neutrino decoupling, which usually occurs around $1$ second 
($T \sim 1 \, MeV$), 
now takes place when $T \sim 50 \, eV$ 
because the weak interactions are so much stronger. 
Thus, the generation of light nuclei 
(of the type described in subsection A above) takes place 
under equilibrium conditions. 
The ratio of the number of protons to the 
number of neutrons in nuclei will be somewhat 
less than one (in the usual standard model it is $7$ to $1$). 
Big Bang Nucleosynethesis also takes place in the presence 
of tau's, muons and electrons and their anti-particles 
and in the presence of many neutral mesons 
because these states are all light. 
It also happens a little earlier:  
from a fraction of a second to one minute 
because the universe expands more quickly and 
cools more rapidly due to the presence 
of these additional light particles. 
In addition, as the temperature drops, 
proton decay becomes important. 

Recombination only takes place after the universe 
has become frigid and if sufficiently stable charged nuclei exist. 
When $T \sim 10^{-7} \, eV$, 
corresponding to when the universe is $10^{15}$ years old,
muons bind to any stable charged nuclei. 
The tau leptons actually decay before having a chance to form atoms.  

When the universe reaches one billion years old, 
it is still possible that galaxies and stars form 
when clouds of nuclei and charged leptons collapse. 
The main difference compared to the usual standard model 
is that the universe remains in a plasma 
for billions of years. 
Also noteworthy is the presence of positions, muons, anti-muons, 
tau's and anti-tau's --  
particles that are absent in the case of the usual standard model.  
It would be interesting to investigate 
the detailed evolution of such a universe, 
but this topic is beyond the goals of the present work.

\medskip

{\bf\large\noindent Acknowledgments}  

This work was supported in part 
by the PSC Board of Higher Education at CUNY and   
by the National Science Foundation under the grant  
(PHY-9420615). 
I thank V.\,P.\,Nair and Al Mueller for brief discussions.

%\pagebreak
\bigskip

\def\NPB#1#2#3{ {Nucl.{\,}Phys.{\,}}{\bf B{#1}} ({#3}) {#2}} 
\def\PLB#1#2#3{ {Phys.{\,}Lett.{\,}}{\bf {#1}B} ({#3}) {#2}} 
\def\PRL#1#2#3{ {Phys.{\,}Rev.{\,}Lett.{\,}}{\bf  {#1}} ({#3}) {#2}} 
\def\PRD#1#2#3{ {Phys.{\,}Rev.{\,}}{\bf D{#1}} ({#3}) {#2}} 
\def\PR#1#2#3{ {Phys.{\,}Rep.{\,}}{\bf {#1}} ({#3}) {#2}} 
\def\OPR#1#2#3{ {Phys.{\,}Rev.{\,}}{\bf {#1}} ({#3}) {#2}} 
\def\NC#1#2#3{ {Nuovo Cimento{\,}}{\bf {#1}} ({#3}) {#2}}

\vfill\eject 
\end{document}